\documentstyle[pre,preprint,epsfig,eqsecnum,aps]{revtex}
\begin{document}
\draft
\tightenlines
%\wideabs{
%\preprint{HEP/123-qed}
\title{Instability of scale-free networks under node-breaking avalanches}
\author{Y. Moreno$^1$\cite{byline}, J.B.G\'{o}mez$^2$, A. F. Pacheco$^2$}
\address{$^1$ The Abdus Salam International Centre for
Theoretical Physics,\\ Condensed Matter Group, P.O. Box 586, Trieste,
I-34014, Italy.\\ $^2$ Facultad de Ciencias, Universidad de Zaragoza,
50009 Zaragoza, Spain.}
\date{\today}
\maketitle
%\widetext
\begin{abstract}
The instability introduced in a large scale-free network by the
triggering of node-breaking avalanches is analyzed using the
fiber-bundle model as conceptual framework. We found, by measuring
the size of the giant component, the avalanche size distribution
and other quantities, the existence of an abrupt transition. This
test of strength for complex networks like Internet is more
stringent than others recently considered like the random removal
of nodes, analyzed within the framework of percolation theory.
Finally, we discuss the possible implications of our results and
their relevance in forecasting cascading failures in scale-free
networks.
\end{abstract}
%\begin{multicols}{2}
%\narrowtext
%}
%\section{Introduction}
%\label{intro}

The recent months have witnessed a great effort devoted to the
unraveling of the properties of complex networks
\cite{ama00,str01}. These properties include the rules followed in
the process of formation of the net, the resulting connectivity
distribution of the networks and their robustness under
unfavorable circumstances like the presence of acting damaging
agents, etc \cite{wat98,bar99,alb00}. This general interest is
mainly due to the fact that the subject of complex networks has a
considerable impact on many branches of science and technology and
also in sociology \cite{will00,bha99,jeo00,alb99,fal99,ves01}. An
important observation has been to recognize that some significant
networks have a scale-free connectivity distribution which means
that the number of links emerging from one node statistically
follows a power-law distribution $P_k\sim k^{-\gamma}$. In
particular, it has also been noted that the Internet belongs to
this class of networks with several studies reporting an exponent
$\gamma=2.2\pm 0.1$ \cite{fal99,romu01}.

In this Letter, we explore the robustness of large scale-free
networks in a scenario in which the failure of a node may trigger
the subsequent failure of its neighbors. Two nodes are considered
as nearest neighbors if they are connected by a direct link. It is
thus clear that the possibility of having $2^{nd}$-step failures
can in turn induce $3^{rd}$-step failures, etc and thus a breaking
avalanche can be generated. This idea is not just an entertainment
of theoreticians; as is now well-known, on 10 August 1996, a fault
in two power lines in Oregon led, through an avalanche of
cascades, to a large-scale blackout in the U.S. and Canada.

To implement our model, we will use the framework of the so-called
fiber-bundle models (FBM) \cite{h90,sor00,us00}. In FBM a set of
$N\gg 1$ fibers (elements) is located on the sites of a -usually
regular- lattice, and one assigns to each element a random
strength threshold sampled from a given probability distribution
(the Weibull distribution is frequently used for this purpose).
Then, the set is loaded by uniformly elevating the weight acting
on each element up to a certain value, $\sigma$. All the elements
whose thresholds are lower than $\sigma$ fail in the first
instance. The individual load carried by each of the broken
elements is equally transferred to their surviving nearest
neighbors and therefore the rupture of an element may induce
secondary failures which in turn may trigger tertiary failures and
so on. These systems are usually conservative which means that
when equilibrium is finally attained, i.e., when there occur no
more casualties, the total weight borne by the $N_s$ surviving
elements is equal to $N_s\sigma$. The translation of the FBM
terminology to that of complex networks studied here is simple. A
fiber may be viewed as a node of the net, the directions of the
load transfers are now the links of the net connecting the nodes
to each other and the load can represent, for instance, the
intensity of electric current flowing into the nodes of the
network, or the viral pressure in Internet, etc, i.e. any
magnitude or agent able to surpass a security threshold of an
individual node, break it, and then provoke an increase of
potential damage in the neighborhood of that node.

Recently, several scenarios of instability have been considered
for complex networks. In particular for scale-free networks the
random removal of nodes in a fixed proportion and its impact on
the global connectivity and functionality of the network has been
explored \cite{alb00} with tranquilizing conclusions: networks
with $\gamma<3$ are completely stable. In this scenario, akin to
percolation, the removal of a node never leads to cascades as it
does in our model. This implies that the test of strength analyzed
here is more stringent and thus offers more guarantees for the
security of networks with power law connectivity distributions.

To study the instability of scale-free networks under
node-breaking avalanches, we first construct a network using the
Barab\'{a}si-Albert (BA) algorithm \cite{bar99}. This is a
stochastic growth model in which, at each time step, a new node is
added and attached preferentially to the already existing ones. At
the initial state, we start from a small number $m_0$ of
disconnected nodes and the network grows by adding one new node at
each time step. This new node is connected preferentially to $m$
old nodes with a probability that depends on the node
connectivities $k_i$ through the relation $\Pi(k_i)=k_i/\sum_j
k_j$. By iterating this scheme a sufficient number of times, a
network consisting of $N$ nodes with connectivity distribution
$P_k=2m^2k^{-3}$ and average connectivity $\langle k\rangle=2m$
develops \cite{bar99}. It is worth noting that the preferential
attachment rule introduced in the BA model accounts for the
rich-gets-richer feature observed in many real complex networks.
Besides, the BA model has been recently improved by adding several
new ingredients in order to account for connectivity distributions
with exponents $2<\gamma<3$. The main feature of the scale-free
networks is that each node has a statistically significant
probability of having a very large number of connections compared
to the average connectivity of the network $\langle k\rangle$.
This is not the case for other complex networks \cite{ama00,str01}
where the connectivity distribution is peaked at $\langle
k\rangle$ and decays exponentially fast for $k\gg\langle k\rangle$
and $k\ll\langle k\rangle$. Thus we expect that working with
$\gamma=3$ does not alter the results one would obtain for the
general case $2<\gamma\leq 3$, a fact confirmed by preliminary
numerical simulations in more general SF networks \cite{doro01}.

Let us now assume that the scale-free network previously generated
is exposed to an external pressure or force $F$ and that each node
of the network represents an individual element able to support a
finite amount of ``load'' $\sigma$. As noted before this could be
seen as a system where the individual elements are continuously
subjected to external agents able to affect their functionality if
they overcome a given security threshold. We will also assume that
in the initial state this external force is equally distributed
among all the nodes in the network so that each element bears a
load $\sigma=F/N$. Furthermore, we assign to each node a
statistically distributed security threshold $\sigma_{i_{th}}$
($1\leq i\leq N$) taken from a probability distribution. If the
load acting on a node surpasses its threshold, the node fails and
its load is equally transferred to the non-failed nodes directly
linked to it. This may provoke other nodes to collapse and the
cascades of failure events last until all the sound elements in
the network bear a load lower than their threshold values. In
order to assign the threshold values we will use the Weibull
distribution $P(\sigma_{i_{th}})=1-e^{-(\sigma_{i_{th}})^{\rho}}$,
where $\rho$ is the so-called Weibull index which controls the
degree of threshold disorder in the system (the bigger the value
of $\rho$, the narrower the range of threshold values). This
allows us to compare the stability of systems having different
levels of heterogeneity in their security threshold distribution.

The threshold rule introduced above has been used for many years to
study a wide class of nonequilibrium phenomena
\cite{sor00,marro,jen98}. However, they have been extensively studied mainly
for {\em regular} lattices. In the present model, the cascading
process not only depends on the thresholds of the elements but also on
the distribution of the neighbors, i.e., a casualty at a node is
determined by both the threshold of that node and the number (and
state) of nodes directly linked to it.

We have performed large scale numerical simulations of the
cascading process produced by repeatedly applying the rules stated
above. In the initial state all the elements that form the network
are subjected to a small individual load $\sigma$. If that load is
bigger than one (or several) threshold(s), a cascading process
could start that lasts until the system arrives at a new
equilibrium state where all the nodes support a load lower than
their security thresholds.  Then, the complete process is repeated
again by imposing in the initial state (with all the elements in
the non-failed state) a bigger load and applying the same failure
rules. After each cascading event, the damage to the system
increases, which affects both the properties of the network and
its functionality. Each simulation is performed many times to
average over the security threshold distribution and in the end a
kind of phase diagram for each magnitude characterizing the final
damaged state of the system is obtained.

\begin{figure}[t]
\begin{center}
\epsfig{file=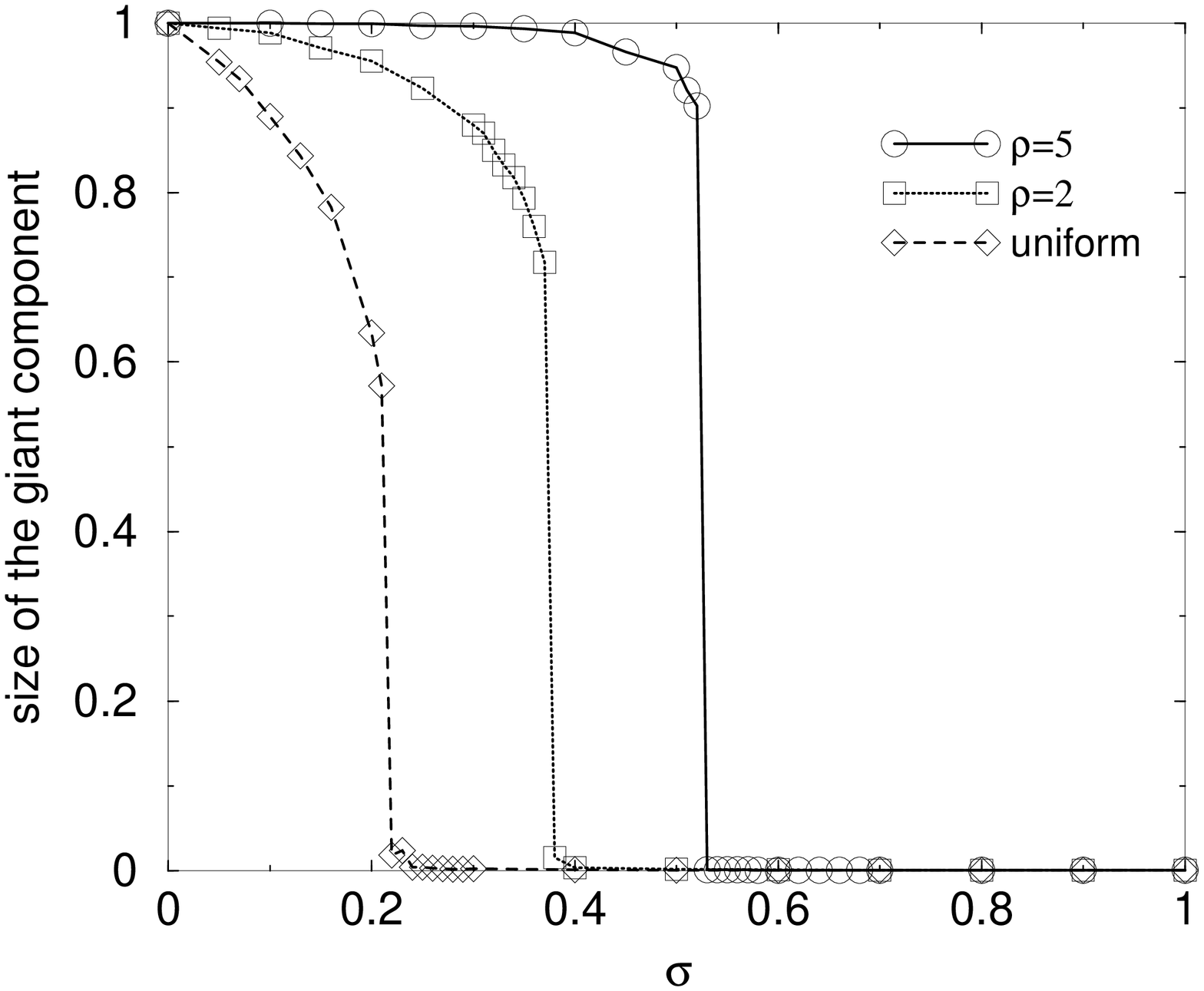,width=8.5cm,angle=-90,clip=1}
\end{center}
\caption{ Size of the giant component for a BA network consisting of
$10^5$ nodes as a function of the load imposed over the system (in
dimensionless units). The
values of $\rho=5$ and $\rho=2$ correspond to two different
levels of heterogeneity for the security thresholds of the nodes. As
the threshold distribution is more homogeneous, the critical point
shifts righwards; however, the precursory activity is less intensive and
the collapse of the network more catastrophic.}
\label{figure1}
\end{figure}

The results obtained indicate that the system has an abrupt
transition in its connectivity. Because of the similitude between
this behavior and the one found in fracture systems on regular
networks, we would call it a critical point. Figure \ref{figure1}
shows the size of the giant component in a BA network composed by
$N=10^5$ nodes as a function of the control parameter $\sigma$ for
a Weibull distribution with two different values of $\rho$ and for
a uniform distribution of thresholds. The size of the giant
component, which plays the role of an order parameter, is defined
as the total number of intact nodes remaining in the largest
component of the network after the cascading events divided by the
system size $N$. We performed a depth-first search for the largest
component \cite{lan00} and averaged the results after many
realizations to get a stable mean value of this component's size.
As can be seen from the figure, for small values of the weight
imposed over the system, the network remains almost intact and the
giant component size is still large enough to ensure the network's
functionality. However, as the load is increased, the cascading
failure begins to reach more and more nodes up to the critical
point where the size of the giant component suddenly falls close
to zero provoking the rupture of the system in many small parts
losing its properties as a functional network.

Figure \ref{figure1} also shows that the network with the highest
degree of homogeneity in the threshold distribution is more
resilient to breakdown. This behavior is the opposite to what is
seen in regular networks. Although the critical load at which the
network loses its functionality shifts to the right as the level
of homogeneity in the thresholds is increased, the precursory
activity is less intensive and so the final breakdown of the
network arises more abruptly and catastrophically, without
previous warning. This result agrees with what is seen in fracture
processes. Obviously, from a practical point of view, this is as
unwanted as having a low critical value that makes the network
very unstable. An intermediate value could satisfy both criteria;
one is to have a robust network and the other to guarantee that
the failure of the network is preceded by an important precursory
activity which helps to foresee the cascades and the imminent
collapse.

\begin{figure}[t]
\begin{center}
\epsfig{file=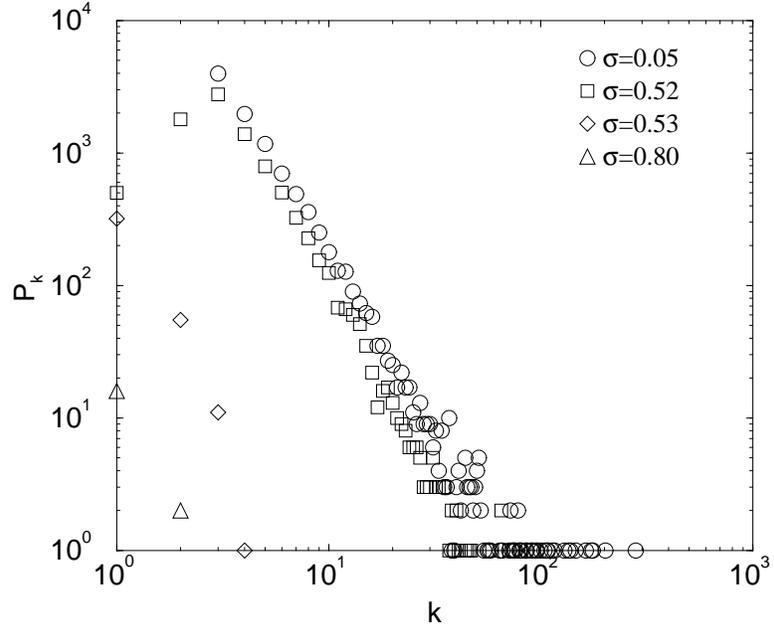,width=8.5cm,angle=-90,clip=1}
\end{center}
\caption{ Connectivity distribution $P_k$ of the network after the
cascading process for several values of the load imposed on the
system. The model's parameters are: $N=10^5$ nodes and $\rho=5$.
The critical load is $\sigma_c=0.52$. The connectivity
distribution remains almost unaltered until the system reaches its
critical point where it is split into many parts (see the
distribution for $\sigma=0.53$, just above $\sigma_c$).}
\label{figure2}
\end{figure}

Another way to shed light on the cascading process is to inspect
the change in the topology of the network as the control parameter
varies. The transition from a functional network to the fragmented
one is illustrated by the simulation results shown in Fig.\
\ref{figure2}, where we plot the probability that a node has
connectivity $k$ when the system has reached the final equilibrium
state. For very low damage pressures (for instance, $\sigma=0.05$
in the figure), the topology of the network remains unchanged.
Right after the critical point ($\sigma=0.52$ in Fig.\
\ref{figure2}), the system loses its properties as an scale-free
network and the whole system becomes disconnected, the largest
connectivity of the nodes being of about 4 links.

\begin{figure}[t]
\begin{center}
\epsfig{file=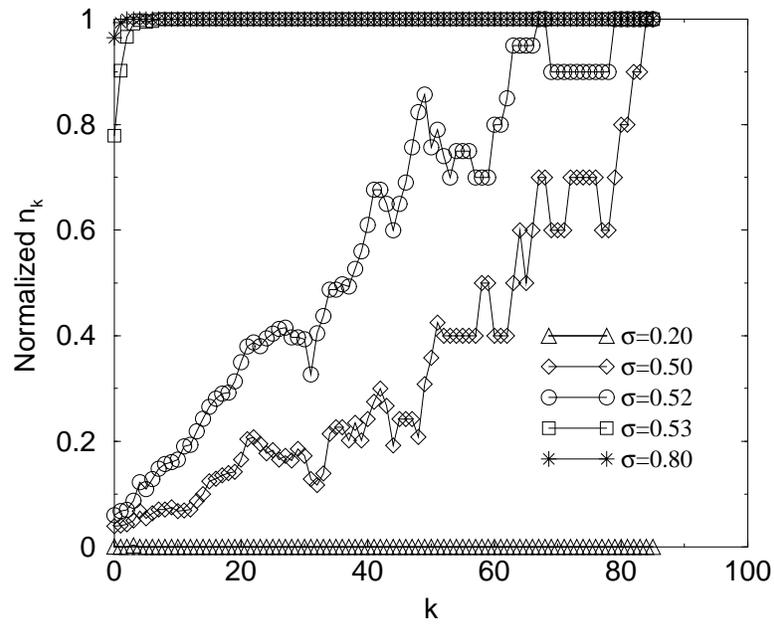,width=8.5cm,angle=-90,clip=1}
\end{center}
\caption{Normalized fraction of broken nodes with connectivity $k$
as a function of their connectivity. The existence of a critical
point is again clearly manifested. Note that at the critical point
there remain in the network a few hubs that ensure the presence of
a giant component in the system. The model parameters are the same
as Fig.\ \ref{figure2}.} \label{figure3}
\end{figure}

We have also monitored the nodes that collapse more frequently.  This
can be observed in Fig.\ \ref{figure3} where we have represented the
fraction of broken nodes $n_k$ with respect to their connectivities
for the same parameters of Fig.\ \ref{figure2}. The existence of a
critical point is again clearly demonstrated. As the load supported by
the system approaches the breakdown point from below, the nodes with
higher connectivities are the most affected by the cascading process
although the system still conserves a few hubs that make possible the
existence of a large giant component that keeps a significant fraction
of nodes interconnected. Right after the critical point only nodes
forming isolated clusters remain, none of them having a large
connectivity. While the fragility of scale-free networks with respect
to the removal of highly connected nodes has been recently documented
by several authors \cite{alb00}, it is worth noting that the above
result, although pointing in the same direction, is obtained as a
consequence of the model's rules instead of being imposed from the
outside, i.e., the system is subjected to an external force and it
evolves according to a simple threshold rule in contrast to other
models \cite{alb00}where the removal of hubs is directed.

\begin{figure}[t]
\begin{center}
\epsfig{file=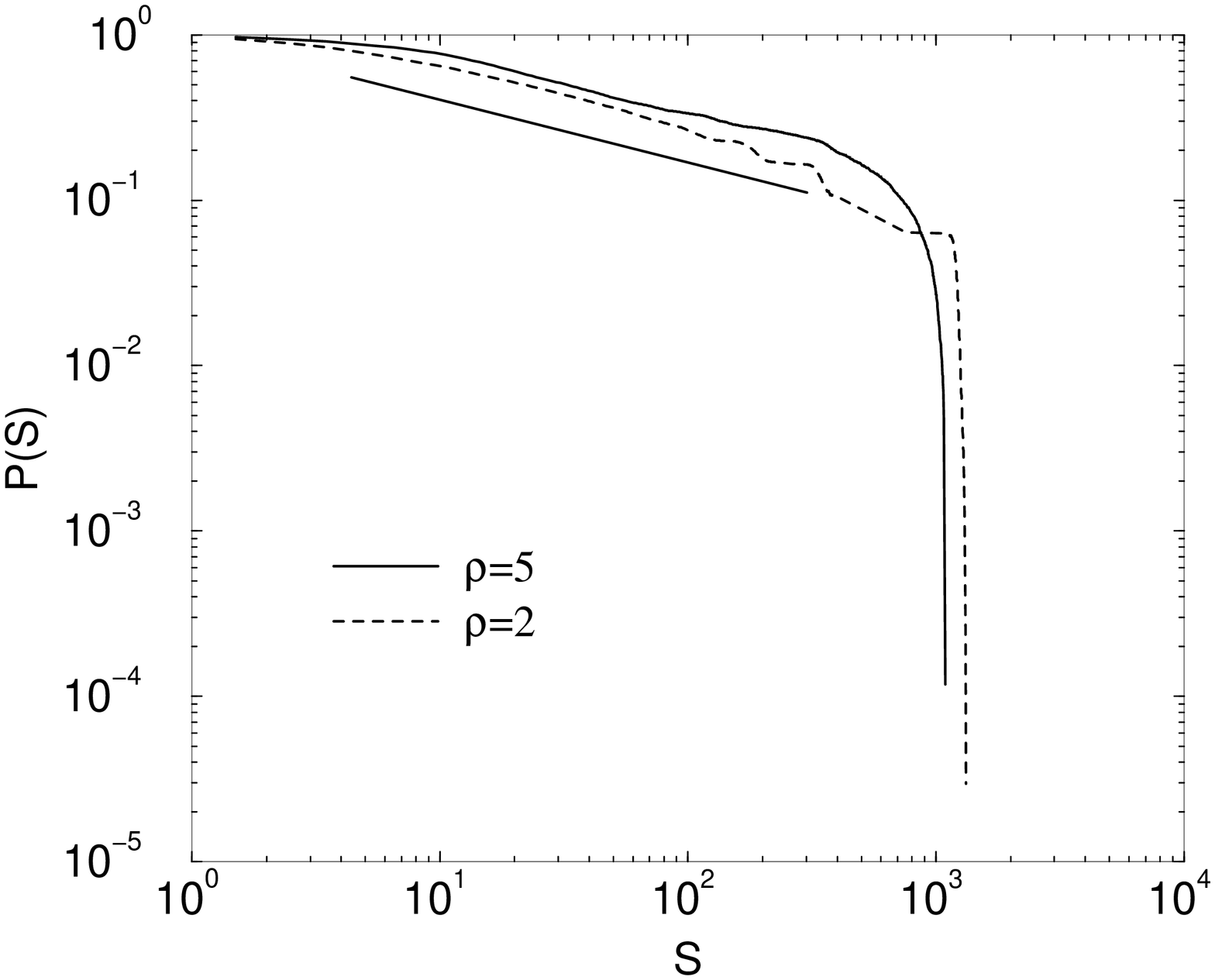,width=8.5cm,angle=-90,clip=1}
\end{center}
\caption{Cumulative avalanche size distributions at the critical
point for $\rho=5$ and $\rho=2$. The system consists of $N=10^5$
elements and each curve has been obtained after averaging over
$10^4$ different realizations of the disorder. A full line with a
slope of 0.12 has been drawn for comparison. Above the critical
point the probability distribution function splits into two parts
and is characterized by the excess of large avalanches. For very
homogeneous threshold distributions the avalanche size
distribution becomes strongly peaked.} \label{figure4}
\end{figure}

Finally, we have characterized the cascading process itself by
measuring the avalanche distribution. The size $s$ of an avalanche is
defined as the total number of nodes that break simultaneously.  The
cumulative distributions $P(s)$ of avalanche sizes for a network
composed of $10^5$ nodes have been represented in Fig.\ \ref{figure4}
for two different values of the threshold disorder parameter. In both
cases, the avalanche size distribution was measured at the critical
points which are in these cases $\sigma_c=0.52$ and $\sigma_c=0.37$
for $\rho=5$ and $\rho=2$ respectively. The distributions can be
fitted, for low values of $\rho$, by a power law of the form $P(s)\sim
s^{-\tau}$, giving an exponent for the probability distributions of
$\tau+1=1.12\pm 0.03$. As the threshold distribution gets more
homogeneous (i.e., bigger $\rho$), the avalanche size distribution
becomes a strongly peaked function around a large mean value. This
occurs because, as we move to the region of large critical points
values (increasing $\rho$), there is a very poor precursory activity
and almost all the nodes forming the network breakdown in only one
time step. Above the critical point, the probability distribution of
avalanches splits into two parts and is characterized by the excess of
large avalanches signaling that we are in the supercritical
region. Without providing a figure, we also want to report that in
scale-free networks the value of the critical stress, for the same
$\rho$, is independent of the size of the system. This differs from
what it is obtained in regular lattices simulating non-mean field
fracture models: there a bigger system implies a lower critical stress
\cite{us00}.

During the completion of this work, we have become aware of a similar
study by Watts \cite{watts00}. He presents an interesting analytical
approach to this kind of problems and solves the small-world
case. Both models and the results reported, however, are different. In
our model the initial distribution of intact and broken nodes after
the first casualty have no contrains as in Watts' model. Besides, and
more important, the relationship between heterogeneity level, the
stability of the system and the precursory activity is not accounted
for in \cite{watts00}. We refer the reader to \cite{watts00} for more
details. 

To sum up, we have introduced a model that accounts for the cascading
events observed in many complex networks. By imposing an external
pressure over the system, several magnitudes have been recorded and
the system has been shown to exhibit a sort of critical point. The
results point out that in order to prevent the breakdown of scale-free
networks, one has to find an optimal criterion that takes into account
two factors: the robustness of the system itself under repeated
failures and the possibility of knowing in advance that the collapse
of the system is approaching.

Y.\ M.\ thanks A. Vespignani for many useful and stimulating
discussions. This work was partially supported by the Spanish DGICYT
Project PB98-1594.

\newpage

\newpage

\newpage

\newpage

%\end{multicols}

\begin{references}
\bibitem[*]{byline} e-mail:yamir@ictp.trieste.it

\bibitem{ama00} L. A. N. Amaral, A. Scala, M. Brath\'{e}l\'{e}mi, and
H. E. Stanley, {\em Proc. Nat. Acad. Sci.} {\bf 97}, 11149 (2000).

\bibitem{str01} S. H. Strogatz, {\em Nature} {\bf
410}, 268 (2001).

\bibitem{wat98} D. J. Watts and S. H. Strogatz, {\em Nature} {\bf
393}, 440 (1998).

\bibitem{bar99} A.-L. Barab\'{a}si, and R. Albert, {\em Science} {\bf
286}, 509 (1999); A.-L. Barab\'{a}si, R. Albert, and H. Jeong, {\em
Physica A} {\bf 272}, 173 (1999).

\bibitem{alb00} R. Albert, H. Jeong, and A.-L. Barab\'{a}si, {\em Nature} {\bf
406}, 378 (2000); R. Cohen, K. Erez, D. ben-Avraham, and S. Havlin,
{\em Phys. Rev. Lett.} {\bf 85}, 4626 (2000); D. S. Callaway,
M. E. J. Newman, S. H. Strogatz, and D. J. Watts, {\em
Phys. Rev. Lett.} {\bf 85}, 5468 (2000).

\bibitem{will00} R. J. Williams and N. D. Martinez, {\em Nature} {\bf
404}, 180 (2000).

\bibitem{bha99} U. S. Bhalla and R. Iyengar, {\em Science} {\bf
283}, 381 (1999).

\bibitem{jeo00} H. Jeong, B. Tombor, R. Albert, Z. N. Oltavi, and
A.-L. Barab\'{a}si, {\em Nature} {\bf 407}, 651 (2000).

\bibitem{alb99} R. Albert, H. Jeong, and A.-L. Barab\'{a}si, {\em Nature} {\bf
401}, 130 (1999).

\bibitem{fal99} M. Faloutsos, P. Faloutsos, and C. Faloutsos, {\em
Com. Comm. Rev.} {\bf 29}, 251 (1999).

\bibitem{ves01} R. Pastor-Satorras, and A. Vespignani, {\em
Phys. Rev. Lett.} {\bf 86}, 3200 (2001); R. Pastor-Satorras, and
A. Vespignani, {\em Phys. Rev. E} (in press) [preprint cond-mat/0102028].

\bibitem{romu01} R.Pastor-Satorras, A. V\'{a}zquez, and A. Vespignani,
preprint cond-mat/0105161.

\bibitem{h90} {\sl Statistical Models for the Fracture of Disordered
Media\/}. Editors, H.J. Herrman and S. Roux, North Holland (1990), and
references therein.

\bibitem{sor00} {\em Critical Phenomena in Natural Sciences},
D. Sornette, Springer-Verlag, Berlin (2000), and references therein.

\bibitem{us00} Y. Moreno, J. B. G\'{o}mez, A. F. Pacheco,{\em
Phys. Rev. Lett.} {\bf 85}, 2865 (2000); S. Zapperi, P. Ray,
H. E. Stanley, A. Vespignani, {\em Phys. Rev. Lett.} {\bf 78}, 1408
(1997); W.I. Newman, A.M. Gabrielov, T.A. Durand, S.L. Phoenix and
D.L. Turcotte, {\em Physica D}{\bf 77}, 200 (1994);
M. V\'{a}zquez-Prada, J. B. G\'{o}mez, Y. Moreno, A. F.  Pacheco, {\em
Phys. Rev. E} {\bf 60}, 2581 (1999); F. Kun, S. Zapperi,
H. J. Herrmann, {\em Eur. Phys. J.} {\bf B17}, 269 (2000).

\bibitem{doro01} S. N. Dorogovtsev, J. F. F. Mendes, and
A. N. Samukin, preprint cond-mat/0106141.

\bibitem{marro} J. Marro and R. Dickman, {\em Nonequilibrium phase
transitions in lattice models} (Cambridge University Press, Cambridge,
1999).

\bibitem{jen98} H. J. Jensen, {\it Self-Organized Criticality}
(Cambridge University Press, Cambridge, England, 1998) and references
therein; D. Stauffer, and A. Aharony, {\it Introduction to Percolation
Theory} (2nd Edition, Taylor and Francis, London 1994). 

\bibitem{lan00} D. Landau, K. Binder, {\em A guide to Monte Carlo
simulations in statistical physics} (Cambridge University Press,
Cambridge, 2000).

\bibitem{watts00} D. J. Watts, {\em Proc. Nat. Acad. Sci.}, in
press. Also available at \\
http://netec.mcc.ac.uk/WoPEc/data/Papers/wopsafiwp00-12-062.html .

\end{references}
\end{document}